\documentclass[twocolumn, superscriptaddress, prb, showpacs, footinbib]{revtex4-1}
\usepackage{graphicx}

\begin{document}

\title{When metal organic frameworks turn into linear magnets}

\author{Pieremanuele Canepa}
\affiliation{Department of Physics, Wake Forest University,
Winston-Salem, NC 27109, USA}

\author{Yves J. Chabal}
\affiliation{Department of Materials Science and Engineering, University
of Texas at Dallas, TX 75080, USA.}

\author{Timo Thonhauser}
\email{thonhauser@wfu.edu}
\affiliation{Department of Physics, Wake Forest University,
Winston-Salem, NC 27109, USA}

\date{\today}


\begin{abstract} 
We investigate the existence of linear magnetism in the metal organic
framework materials MOF-74-Fe, MOF-74-Co, and MOF-74-Ni, using
first-principles density functional theory. MOF-74 displays regular
quasi-linear chains of open-shell transition metal atoms, which are well
separated. Our results show that within these chains---for all three
materials---ferromagnetic coupling of significant strength occurs. In
addition, the coupling in-between chains is at least one order of
magnitude smaller, making these materials almost perfect 1D magnets at
low temperature. The inter-chain coupling is found to be
anti-ferromagnetic, in agreement with experiments. While some quasi-1D
materials exist that exhibit linear magnetism---mostly complex oxides,
polymers, and a few other rare materials---they are typically very
difficult to synthesize. The significance of our finding is that MOF-74
is very easy to synthesize and it is likely the simplest realization of
the 1D Ising model in nature. MOF-74 could thus be used in future
experiments to study 1D magnetism at low temperature.
\end{abstract}

\maketitle

The continued quest for the development of non-volatile memories and spintronic
devices of smaller sizes requires the full comprehension of finite-size
effects.  To this end, over the last decade, exotic magnetic properties have
received much attention in experimental and theoretical
studies.\cite{Pratzer01, Clerac02, Gambardella02, Johannes06, Beobide07,
Kimura08, Kimura08a, Kurmoo09, Sun10, Kawasaki11, Simon11,
Toma12,Stone12,Zhang10} Considerable emphasis has been given to the synthesis
and prediction of materials showing mono-dimensional magnetism,\cite{Pratzer01,
Clerac02, Gambardella02, Johannes06, Beobide07, Kimura08, Kimura08a, Kurmoo09,
Sun10, Kawasaki11, Simon11, Toma12,Stone12,Zhang10} also referred to as
\emph{1D} or \emph{linear} magnetism.  While 1D magnetism can be explained with
the well-understood Ising model (dating back to 1925),\cite{Blundell01} a
satisfactory physical realization of this model in simple materials has not yet
been found and 1D magnetism is only observed in a few---often
difficult\cite{Sun10,Stone12,Kurmoo09} or dangerous\cite{Zhang10} to
synthesize---synthetic inorganic materials and polymers. Although, for example,
CrSb$_2$ is one of the few materials that shows naturally 1D
anti-ferromagnetism, this property remains difficult to control and
tune.\cite{Stone12} In fact, theory has shown that strong spin fluctuations
induce ferromagnetic disorder of 1D-spin arrays at any temperature, independent
of the extent of exchange interactions between neighboring
spins.\cite{Mermin66, Blundell01} Thus, progress in the field of 1D magnetism
crucially depends on the availability of currently missing simple-to-synthesize
model systems and materials.

The main difficulty in engineering good model systems exhibiting 1D magnetism
are:\cite{Clerac02, Sun10} (i) to find materials that have quasi-1D chains of
spins with significant interactions and large magnetic anisotropy, (ii) to find
materials with a large ratio between intra and inter-chain magnetic
interactions, (iii) to find materials in which ferromagnetism is preserved at
``reasonable'' low temperatures, (iv) to find materials with very few
impurities, which tend to destroy ferromagnetism, and finally (v) to find
materials that are simple, safe, inexpensive to synthesize, and where linear
magnetism is easy to control.  Historically, the engineering of 1D magnetic
materials has followed several routes. Attempts were made using inorganic
materials such as Sr$_2$Cu(PO$_4$)$_2$, Sr$_2$CuO$_3$,\cite{Johannes06} and
BaCo$_2$V$_2$O$_8$,\cite{Kawasaki11} along with non-periodic magnetic clusters
or molecular magnets.\cite{Sun10} Another strategy is the combination of
organic molecules and transition metals (TM) to form regular polymer 1D
magnets.\cite{Clerac02,Kurmoo09, Sun10, Toma12}  The latter strategy offers a
larger degree of freedom due to the high tunability of the diamagnetic organic
separators, which can promote spin localization on the central
TM.\cite{Clerac02, Kurmoo09, Sun10}

\begin{figure}
\begin{center}
\includegraphics[width=0.70\columnwidth]{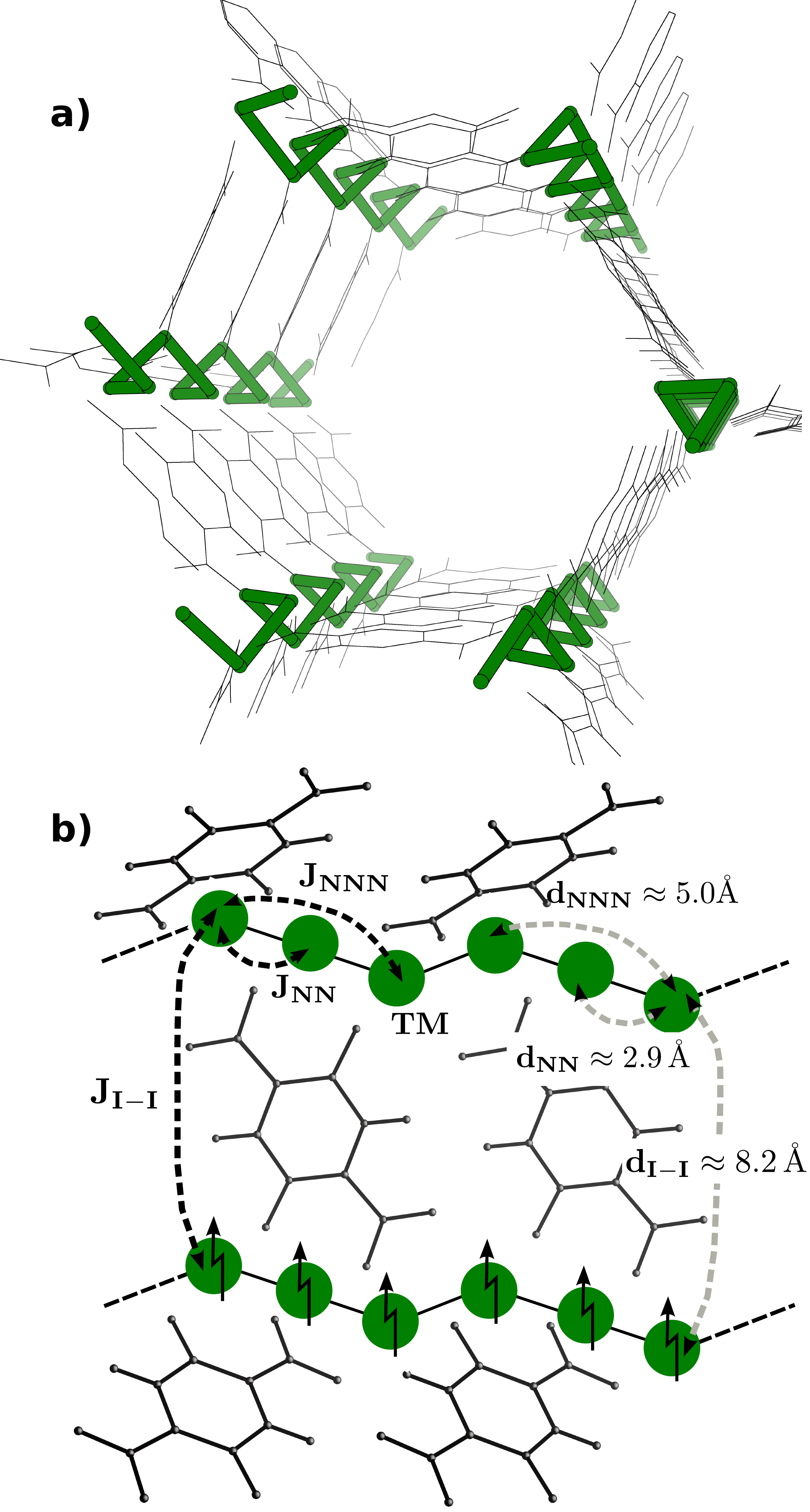}
\end{center}
\vspace{-2ex}
\caption{\label{fig:1_chains} {\bf a)} Frontal view of MOF-74,
helicoidal magnet chains are highlighted in green. {\bf b)} side view of
MOF-74, TM atoms are represented by green balls. $d_{\rm NN}$ and
$d_{\rm NNN}$ are the nearest-neighbor and next-nearest-neighbor
intra-chain distances, while $d_{\rm I-I}$ is the inter-chain distance.
The couplings $J_{\rm NN}$, $J_{\rm NNN}$, and $J_{\rm I-I}$, are
defined in parallel.}
\end{figure}

We propose here that metal organic frameworks (MOFs), a novel class of
nano-porus materials, offer a versatile platform for the realization of 1D
magnets due to their high tailorability and tunability that results from their
discrete molecular building-block nature.\cite{Furukawa07, Murray09, Britt09,
Li09}  For this reason MOFs are already targeted in a large variety of
applications such as gas-separation, gas-sensing, gas-capture, catalysis, and
drug-delivery.\cite{Furukawa07, Murray09, Britt09, Li09, Nijem12,Canepa13} In
particular, in the following, we argue that the structural simplicity, low
cost, and ease of synthesis of MOF-74---together with the already existing
understanding of this material---fulfill the criteria mentioned above and thus
make it an outstanding candidate for studying linear magnetism. Note that
signatures of 1D-ferromagnetism in MOF-74-Co were already observed
experimentally by Dietzel \emph{et al.} in their pioneering work on this
MOF.\cite{Dietzel05}  From Fig.~\ref{fig:1_chains} it is apparent that
MOF-74-TM (with TM = Mn, Fe, Co, Ni, and Cu) can be seen as 1D magnet since it
displays regular pseudo-chains of transition metals aligned along the basal
plane. The helicoidal chains resulting from the atomic-motif of
Fig.~\ref{fig:1_chains}a are interspaced by ``long'' organic linkers,
suggesting that the inter-chain interactions are quenched.  In fact, MOF-74
shows a large structural ratio ($\sim$ 3) between the separation of spins in a
chain compared to chain separation (see Fig.~\ref{fig:1_chains}), establishing
a required condition for the construction of a 1D magnet.

To elucidate the 1D-like magnetic properties exhibited by MOF-74, we study the
three isostructural materials MOF-74-Fe,\cite{Bloch12}
MOF-74-Co,\cite{Dietzel05} and MOF-74-Ni.\cite{Dietzel06} To this end, we use
density functional theory (DFT) with the PBE functional, as implemented in
\textsc{QuantumEspresso}.\cite{Giannozzi09} We employ ultrasoft
pseudopotentials with wave-function and density cutoffs of $680$ eV and $6800$
eV. The pseudopotentials used for the TM (i.e. Fe, Co, and Ni) are also
suitable for spin-orbit calculations including relativistic corrections. The
total energy is sampled with a 2$\times$2$\times$2 \emph{k}-point mesh,
resulting in energy differences converged to within less than 1~meV. Projected
density of states onto selected atomic orbitals are performed on a denser
\emph{k}-point mesh, i.e.\ 4$\times$4$\times$4. The SCF total energy
convergence criterium is 1.4$\times$10$^{-10}$ eV. We need such tight criteria
to be able to accurately sample the delicate energy landscape originating from
different spin arrangements.

All calculations are performed on the experimental structures of
MOF-74-Fe,\cite{Bloch12} MOF-74-Co,\cite{Dietzel05} and
MOF-74-Ni,\cite{Dietzel06} which crystallize in a rhombohedral primitive cell
with 54 atoms and space group $R\overline{3}$.  The calculation of the
intra-chain $J$-coupling constants requires the freedom to have varying spin
directions along a chain. But, the primitive cell of MOF-74-TM contains only 6
TM atoms that all belong to different chains (1 per chain), which does not give
the required freedom. Thus, we construct a supercell extending the unit cell
along the chain direction, such that each unit cell now contains two chains
with 6 TM atoms per chain, and a total of 108 atoms.  Coordinates and relative
lattice constants of the supercells are reported in the Supplementary
Information.

Linear magnetism relies on ferro- or antiferro-magnetism that can only exist if
the TM atoms have a non-negligible magnetic moment. We therefore begin by
analyzing the localization of the magnetic moment on the TM atoms, combining
the projected density of states and the L\"{o}dwin population analysis. The
L\"{o}dwin analysis, similarly to the Mulliken analysis is an intuitive (but
not unique) way of re-partitioning the electron charge density on each atom
(and orbital), by projecting it onto individual orthonormalized atomic
orbitals.\cite{Szabo}  Table~\ref{table:charges} shows the L\"{o}dwin charges,
relative contribution, and magnetic moments of the TM and O atoms in the three
MOF-74 investigated.  The magnetic moments, $\mu$, were computed by integrating
the spin-densities difference ($\rho_{\rm up} -\rho_{\rm down}$) of the
\emph{d-p} orbitals in the valence region of each TM.  Although it is
inadequate to draw decisive conclusions from the charge analysis of
Table~\ref{table:charges}, we observe that oxygen atoms in MOF-74 assume an
interesting covalent nature, having repercussions on the final charges and
magnetic moment of the TM in MOFs. We further confirm the local charge of Co in
MOF-74-Co (+2.49), which was experimentally assigned as 2+.\cite{Dietzel05} It
is also interesting to see that the local charge of Fe in MOF-74-Fe behaves
almost like the metallic case, thus increasing the local magnetic moment. The
experimental magnetic moment for Co is 4.67 $\mu_B$,\cite{Dietzel05} which is
larger than our computed value; a discrepancy connected to the well-known
unphysical delocalization of the electron charge density that is introduced by
the exchange-correlation functional adopted in DFT simulations.\cite{Iberio06}
Note that orbital magnetism\cite{timo_1, timo_2, timo_3}  is not included in
our calculations, as its effect is typically very small.\cite{timo_4}

\begin{table}
\caption{\label{table:charges} MOF-74-TM net atomic charges (in units of the
electronic charge), $Q_{\rm O}$ and $Q_{\rm TM}$, and electron population of
\emph{p} and \emph{d} orbitals on O  and TM atoms, $q_{\rm O}(2p)$ and $q_{\rm
TM}(3d)$, respectively. Magnetic moments, $\mu$, are reported in units of
$\mu_B$.}
\begin{tabular*}{\columnwidth}{@{\extracolsep{\fill}}lccccc@{}}\hline\hline
TM & $Q_{ \rm O}$ &  $q_{\rm O} (2p)$  & $Q _{\rm TM}$ & $q_{\rm TM} (3d)$  & $\mu$\\\hline
Fe & --0.30 &  4.75 & +0.50 & 6.35 & 3.625 \\
Co &  --0.95 & 5.35  &  +2.49 & 5.17 & 3.255\\
Ni & --0.61 & 4.94 &  +1.24 &  8.33 & 1.567\\
\hline \hline
\end{tabular*}
\end{table}

\begin{figure}
\includegraphics[width=\columnwidth]{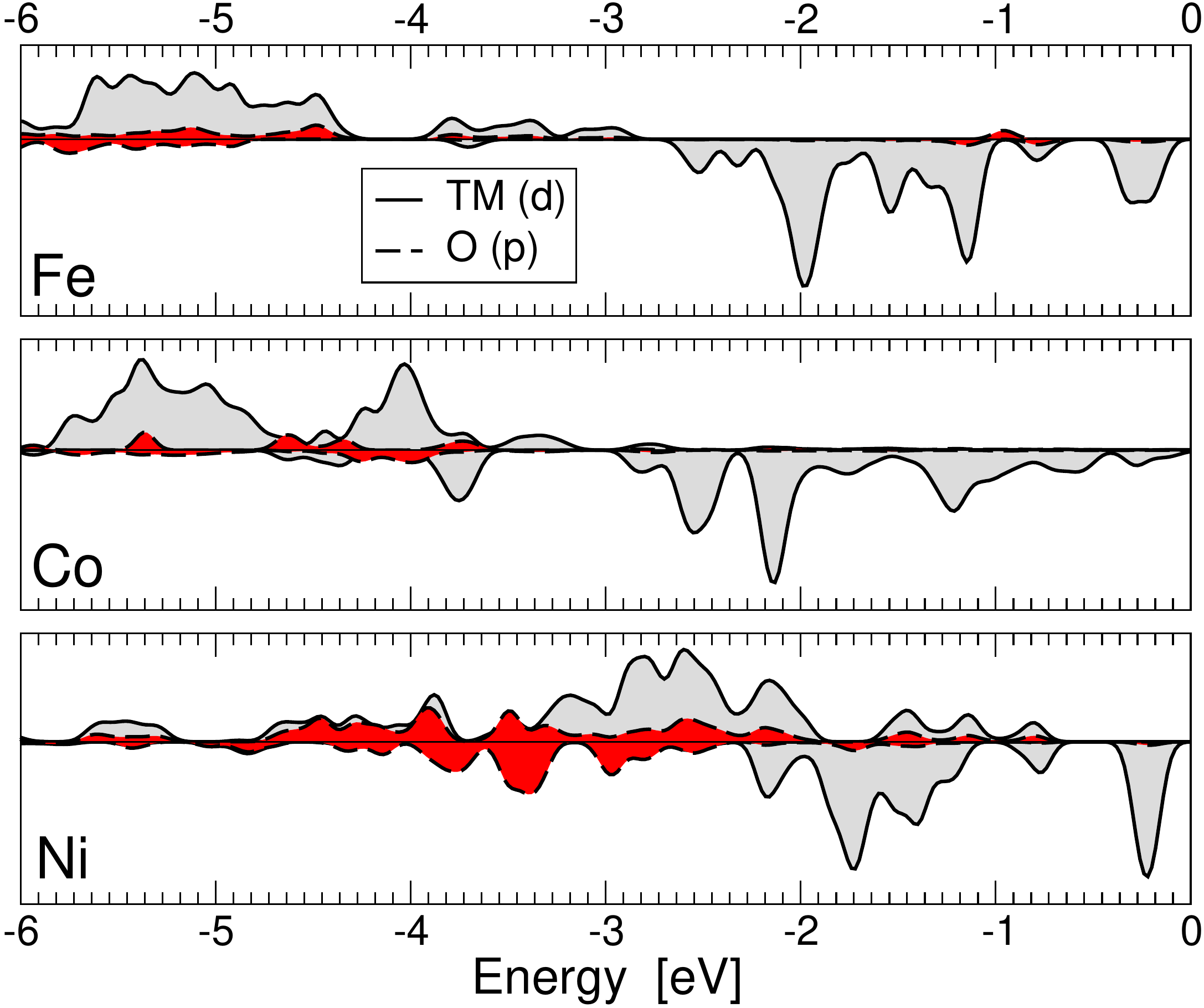}
\caption{\label{fig:2_dos} Projected density of states onto Fe, Co, and
Ni \emph{d}-orbitals (gray) and O \emph{p}-orbitals (red) of the valence
bands of MOF-74-TM.  Energy is given in eV with respect to the top of
the valence band.  Spin-up and spin-down densities are plotted above and
below the zero line of each plot.}
\end{figure}

Figure~\ref{fig:2_dos} shows the density of states of the valence bands
projected onto the \emph{d}-orbitals of the TM atoms and \emph{p}-orbitals of
oxygen atoms (pDOS).  Here we see that some of the electronic charge density of
the TM spills-over (due to orbital hybridization) to the nearest-neighbor oxygen atoms. This diminishes the
local magnetization on spin-carriers and thus their total magnetic moment. Not
surprisingly, the analysis of the pDOS together with the charge analysis
suggests that the magnetization originates from the \emph{d}-electrons
(spin-down, see Fig.~\ref{fig:2_dos}) of the TM atoms. Note that the angle
$\angle_{\rm TM-O-TM}\approx 90^\circ \pm 5^\circ$ does not allow sufficient
overlap between the relevant orbitals enforcing the intrachain ferromagnetism
according to the Goodenough and Kanamori rules.\cite{Goodenough63}  The above
analysis clearly shows how the tunability of the organic linkers in MOFs can be
utilized to increase the spin localization on the TM, and a more involved
explanation can be found in Ref.~\onlinecite{Kurmoo09}. From this analysis we
conclude the existence of localized magnetic moments that can give rise to
ferromagnetic coupling among TM atom chains.

Although we have clarified the existence of chains of spin carriers, we still
need to understand if spin chains are independent of each other (see
Fig.~\ref{fig:1_chains}) in order to produce isolated spin arrays acting as
linear magnets. To this end, a qualitative estimation of the magnetic
independence of spin chains is obtained by performing calculations where each
chain magnetization is assigned a random spatial starting orientation, which
thereafter is free to relax towards the most favorably energetic orientation.
If there is some degree of inter-chain interaction, each chain
spin-magnetization will assume some preferred orientation.  But, our results
show that only a small rearrangement of the spin directions occurs, i.e.\ only
$\pm$~2$^{\circ}$ from the initial directions, supporting the idea that chains
are only weakly coupled. However, a quantitative measurement of such
chain-chain interactions can only be obtained by calculating the inter-chain
$J$-coupling constants, which follows next.

Having established that the TM spin-carriers exhibit a substantial
magnetization that can produce potential ferromagnetic coupling, our
investigation moves to the calculation of the $J$-coupling interactions.
Figure~\ref{fig:1_chains}b shows the magnetic pathways and defines the
following $J$-couplings: The intra-chain $J_{\rm NN}$ and $J_{\rm NNN}$, origin
of the 1D linear magnet properties; and, the unwanted inter-chain $J_{\rm I-I}$
interactions. A complete structural analysis shows that the intra-chain TM-TM
distance falls between 2.8~\AA\ and 3.0~\AA\ for MOF-74-TM, whereas the
intra-chain distance falls between 7.5~\AA\ and 8.8~\AA, giving reason to
believe that the inter-chain $J$-coupling interactions are quenched.  If each
spin magnetization is constrained along the $z$-direction, \cite{footnote} the
coupling interaction, $J_{ij}$, described by the complex Heisenberg-Dirac-van
Vleck Hamiltonian simplifies to the 1D Ising Model \cite{Blundell01} 
\begin{equation} \label{eq:isingham} \hat{H}=-2 \sum _{i,j} ^{n} J_{ij} \,
\hat{S^z _i}\cdot \hat{S^z _j}\;, \end{equation}
where $\hat{S^z_i}$ is the projection of the spin operator along the $z$
direction at site $i$.  Due to the gyromagnetic factor, for the expectation
values of $\hat{S}^z_i$ we use 1/2 of the magnetic moments $\mu$ in
Table~\ref{table:charges}, i.e.\ 0.813 for Fe, 1.628 for Co, and 0.784 for Ni.
We now use DFT to map the real system onto this model Hamiltonian by computing
the energy differences of various ferro- and anti-ferromagnetic spin
configurations, which in turn yields the $J$-couplings.  Our supercell contains
6 independent TM atoms per chain (see Fig.~\ref{fig:1_chains}b), resulting in
$2^6 = 64$ possible different spin configurations, out of which only $16$
combinations are linearly independent and compatible with our periodic boundary
conditions.  The coupling constants $J_{ij}$ are then obtained by solving an
overdetermined system of 16 equations with a least-square fit.
Table~\ref{table:jcoupling} reports our calculated values for the
nearest-neighbor coupling $J_{\rm NN}$, the next-nearest-neighbor coupling
$J_{\rm NNN}$, and the inter-chain coupling $J_{\rm I-I}$ for MOF-74-Fe,
MOF-74-Co, and MOF-74-Ni.  Note that these calculations are a particularly
challenging task requiring high accuracy, as these energy differences are tiny
compared to the total energy of a 108 atom unit cell.

\begin{table} \caption{\label{table:jcoupling} Intra-chain $J$-coupling
constants $J_{\rm NN}$ and $J_{\rm NNN}$ and inter-chain $J_{\rm I-I}$ for
MOF-74-TM in cm$^{-1}$. For clarity we report again the magnetic moment, $\mu$,
in $\mu_B$, from Table~\ref{table:charges}. The standard deviation of $J_{ \rm
NN}$ and $J_{ \rm I-I}$ is not reported because below the accuracy limit.}
\begin{tabular*}{\columnwidth}{@{\extracolsep{\fill}}lcccr@{}}\hline\hline TM &
$\mu$ & $J_{\rm NN}$  & $J_{\rm NNN}$  & $J_{\rm I-I}$\\ \hline Fe & 3.625 &
28.1 $\pm$ 0.4 &  6.0 & --1.2\\ Co & 3.255 & 40.1 $\pm$ 2.9 &  4.9 & --1.9\\ Ni
& 1.567 & 21.0 $\pm$ 3.5 &  6.9 & --1.3\\\hline \hline \end{tabular*}
\end{table}

From Table~\ref{table:jcoupling} we see that the intra-chain $J$-couplings are
larger and more positive than the inter-chain ones, suggesting the existence of
linear ferromagnetism. On the other hand, the interaction among chains is very
small and of anti-ferromagnetic nature.  As expected, longer range $J$-coupling
interactions, such as $J_{\rm NNN}$, are of smaller magnitudes than the
nearest-neighbor interactions and are expected to vanish at increasing
distances.  Although couplings for longer distances are in principle easily
obtainable from Eq.~(\ref{eq:isingham}), such results are not presented here
since they fall below our accuracy limit.  Overall, the trend of the magnetic
constants is maintained between the three TMs.  From our simulations the
computed $J_{\rm NN}$ for MOF-74-Fe seems largely overestimated from the
experimental value of 4.12 cm$^{-1}$, which was extrapolated by fitting
experimental magnetic susceptibility profiles.\cite{Bloch12} On the other hand,
our calculated inter-chain constant for MOF-74-Fe is in excellent agreement
with the experimental result of $-$1.12~cm$^{-1}$.\cite{Bloch12} In summary, we
conclude that the ferromagnetic intra-chain interactions are one order of
magnitude larger than the anti-ferromagnetic inter-chain ones---confirming the
possibility of the existence of 1D-magnetic phenomena at low temperature.

\begin{figure}
\includegraphics[width=1.0\columnwidth]{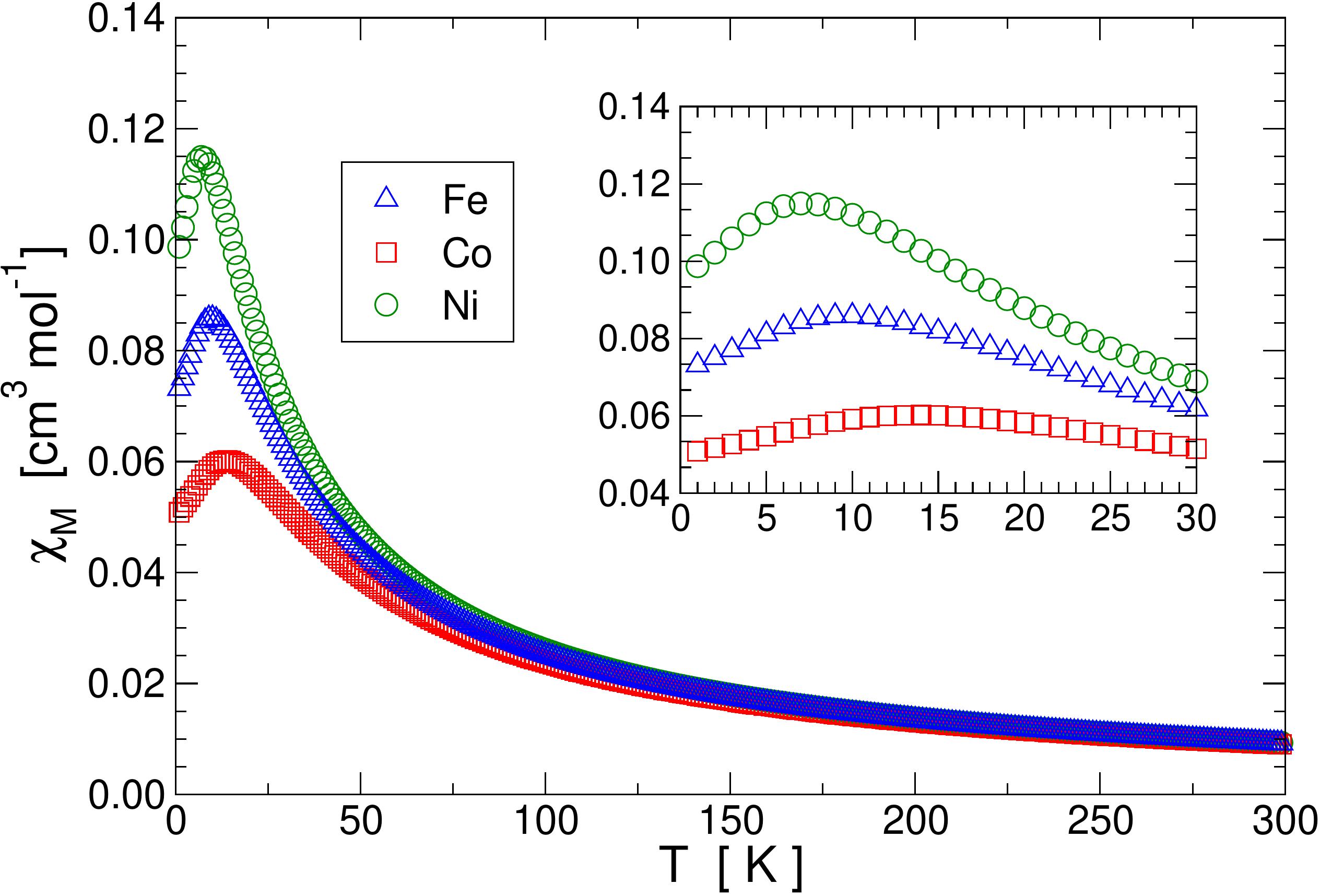}
\caption{ \label{fig:3_magsuce} Computed magnetic susceptibility $\chi
_M$ (in cm$^{3}$ mol$^{-1}$) as a function of temperature $T$ (in K) for
MOF-74-Fe, MOF-74-Co, and MOF-74-Ni. The inset shows an enlargement of
the transition zone.}
\end{figure}

We finally move to calculating the temperature-dependent magnetic
susceptibility $\chi _M$.  Starting from our computed $J$-coupling
constants, we can predict the magnetic susceptibility, $\chi _M$, which
is measurable experimentally.  We use Fisher's model \cite{Fisher64}
\begin{eqnarray}
\label{eq:fisher} 
\chi _M &=& \frac{N g_{\rm iso} ^2 \mu ^2}{12 k_b T} \times
            \frac{1+u(J_{\rm NN})}{1-u(J_{\rm NN})}, \\
u(J_{\rm NN}) &=& \coth \left(\frac{k_b T}{2 J_{\rm NN}}\right) -
                  \left( \frac{2 k_b T}{J_{\rm NN}}\right)\;, \label{eq:fisher2} 
\end{eqnarray} 
where $N$ is the number of atoms in the chain, $g_{\rm iso}$ the g-factor,
$k_b$ the Boltzman constant, and $T$ the temperature.
Figure~\ref{fig:3_magsuce} shows our calculated $\chi _M$ as a function of
temperature for the three MOF-74-TM investigated, using the $J_{\rm
NN}$-coupling constants from Table~\ref{table:jcoupling}.  The transition
temperature corresponding to the phase transition from ferromagnetic order to
anti-ferromagnetic order along the chains is given by the peak position of
$\chi _M$. Obviously, the transition temperature depends on the $J$-coupling
strength: the larger the $J$-coupling constant is, the broader the peak becomes
and the higher the transition temperature.  A similar dependence is found for
the $\chi _M$ magnitude itself, which decreases for increasing $J$-coupling
constant.  For MOF-74-Co, the temperature dependence of $\chi_M$ was measured
experimentally,\cite{Dietzel05} finding a transition temperature of 8--10~K, in
good agreement with our calculated transition temperature of 13~K. The
experimental maximum of the peak is at $\sim$0.17~cm$^3$~mol$^{-1}$, whereas
our calculated maximum is only at $\sim$0.06~cm$^3$~mol$^{-1}$. However, this
discrepancy is a result of the fact that our DFT calculated Co dipole moment of
3.255~$\mu_B$ is too small compared to the experimental one of 4.67~$\mu_B$ (as
mentioned above). \cite{Dietzel05} As can be seen from Eq.~(\ref{eq:fisher})
the dipole moment $\mu$ enters the susceptibility as $\mu^2$. If we simply use
the experimental dipole value, our peak maximum would be at
$\sim$0.13~cm$^3$~mol$^{-1}$, in reasonable agreement with experiment.
Furthermore, note that according to the susceptibility model used, the $J_{\rm
NN}$ coupling constant for MOF-74-Fe has to be larger than the
4.12(6)~cm$^{-1}$ found experimentally through fitting data by Bloch \emph{et
al.};\cite{Bloch12} such a small value results in a transition temperature too
close to 0~K and below the experimental conditions reported in their study
(2--300~K). Equation~(\ref{eq:fisher}) includes only the effect of $J_{\rm
NN}$, making this model quite unsatisfactory.  The effect introduced by
interchain $J_{\rm I-I}$ coupling constant in $\chi _M$ can be reintroduced in
Eq.~(\ref{eq:fisher2}) by replacing $u(J_{\rm NN})$ with $u(\left| J_{\rm NN}/
J_{\rm I-I}\right|)$, with the effect of slightly shifting all curves by $\sim$
--3~K, bringing them in very good agreement with experimental observation. Our
estimated transition temperatures of all three investigated MOFs are clearly
above the liquid He temperature, encouraging further experiments on linear
magnetism phenomena in MOF-74.

In summary, we have explored the existence of linear magnetic phenomena in the
metal organic framework materials MOF-74-Fe, MOF-74-Co, and MOF-74-Ni by using
DFT calculations. Our results provide an understanding of the origins and
magnitude of linear magnetic effects in these materials. We verify the
existence of intra-chain ferromagnetism and quenched anti-ferromagnetic
coupling between chains, large enough to be observed at liquid He temperatures.
The significance of our finding is that MOF-74 is easily synthesized, safe, and
inexpensive. As such, it is likely to be the simplest realization of the 1D
Ising model in nature and has the potential to provide simple means to study
linear magnetism.  Note that the MOFs studied here have not been tailored in
any way to make them good 1D magnets. In view of the high tailorability and
tunability of MOFs, exciting new opportunities open up, where especially
designed linkers and spin-centers decrease the spin-density delocalization  and
maximize the spin moments, resulting in larger $J$-couplings and higher
transition temperatures.

This work was entirely supported by the Department of Energy Grant No.
DE-FG02-08ER46491.



\begin{thebibliography}{99}
\bibitem{footnote} We have considered the more general case of truly
non-collinear spin, but our results show that all three MOF-74 systems indeed
prefer collinear spin arrangements. We thus constrain our calculations here to
spin directions only along the $z$-axis.
\end{thebibliography}
\end{document}